\documentclass{acm_proc_article-sp}
\usepackage{graphics,url}
\usepackage{subfigure}
\usepackage{multirow}

\usepackage[pdfstartview=FitH, bookmarks=false, draft]{hyperref}
\def\sharedaffiliation{%
\end{tabular}
\begin{tabular}{c}}

\newcommand{\tab}{\hspace*{2em}}

\begin{document}
\title{Show Me Your Cookie And I Will Tell You Who You Are}

\numberofauthors{2} %  in this sample file, there are a *total*
\author{
% 1st. author
	\alignauthor Vincent Toubiana \\
          \and		
	\alignauthor Vincent Verdot \\
   \sharedaffiliation
   \affaddr{Alcatel-Lucent Bell Labs --- Application Domain}  \\
   \affaddr{Route de Villejust,  91620 Nozay, France} \\
   \email{[lastname.firstname]@alcatel-lucent.com}
}

\maketitle
\begin{abstract}
\label{abs}
With the success of Web applications, most of our data is now stored on various third-party servers where they are processed to deliver personalized services. Naturally we must be authenticated to access this personal information, but the use of personalized services only restricted by identification could indirectly and silently leak sensitive data.
We analyzed \emph{Google Web Search} access mechanisms and found that the current policy applied to session cookies could be used to retrieve users' personal data. We describe an attack scheme leveraging the search personalization (based on the same \textsc{sid} cookie) to retrieve a part of the victim's click history and even some of her contacts.
We implemented a proof of concept of this attack on Firefox and Chrome Web browsers and conducted an experiment with ten volunteers. Thanks to this prototype we were able to recover up to 80\% of the user's search click history.
\end{abstract}

\section{Introduction}
Over the last few years, Google's core service ``Search'' was enhanced through feature deployments, new UI and display optimization. One major improvement regarding the quality of search results was the personalization of the ranking algorithms. Personalized search results are ranked according to the user's context (\emph{i.e.} localization and language),  profile (search history), social networks and other characteristics extracted from the use of Google's services. Via the \emph{Google Dashboard}, users can view and possibly edit data that was collected by their use of Google services. These data may be very sensitive, so access to this interface naturally requires an authentication.\\
\tab However, while the direct access to users' data is subject to a strict security policy, using personalized services (which may leak this same personal information) is not. Indeed, some Web applications like \emph{Google Search} only verify the (unsecured) user's session to render personalization features. Such a session can be hijacked by simply capturing the corresponding ``\textsc{sid} cookie''. Unlike cookies used to authenticate the user, the \textsc{sid} cookie may be sent cleartext, \emph{i.e.} unprotected. Furthermore, this cookie is sent whenever the user accesses to a service hosted on \url{google.com}, increasing attack opportunities. In this paper we show how the \textsc{sid} cookie could be misused by an attacker, providing ungranted access to Google Search personalized results and history.\\
\tab We study an information leakage attack that exploits the current Google's access policy regarding personalized services (\emph{i.e.} unauthenticated access). More specifically, we hijack a \textsc{sid} cookie to circumvent Google protection and access a user's personal data who --- possibly forced by the attacker --- transmitted her cookie in clear text. We emphasize the risk of using unauthenticated personalized services over a shared network with the following contributions.
\begin{enumerate}
\item The description of an \emph{information leakage} attack that uses the unprotected \textsc{sid} cookie to retrieve the victim's visited search results and a list of her contacts.
\item A proof of concept, based on the browser extension ``Firesheep'', and the tool we used to evaluate the impact of the \emph{information leakage}.
\end{enumerate}
The remaining of this paper is organized as follows. Section \ref{google services and cookies} describes Google's architecture and services relevant to the understanding of the proposed attacks. Section \ref{retrieving data} presents the information leakage attack using the \textsc{sid} cookie to retrieves user's click history and contacts. An implementation of this attack is proposed in Section \ref{implem} along with statistics showing how seriously a Google account's click history could be compromised. Finally section 5 discusses measures that should be deployed to counter this attack and section \ref{conclusion} concludes this paper.

\section{Google Services and Cookies}
\label{google services and cookies}
Google provides more than twenty different services, covering most of people needs over the Web. With a single Google account, a user can access to all these applications even those hosted on different domains (\emph{e.g.} YouTube or Blogger). A couple of cookies are used to help users navigating smoothly between the Google services.\\
\subsection{Google.com cookies}
\tab At least three cookies are systematically sent by the user's browser to Google servers when accessing a service under the \url{google.com} domain.
\begin{itemize}
\item \textsc{pref}: this cookie carries the preferences for the browser currently accessing the service. These preferences refer to the interface, the language, the number of results returned by a search, etc. The \textsc{pref} cookie is attached to a browser and not bound to a specific user account. 
\item \textsc{sid}: this non-secured session cookie is transmitted to Google servers to identify the user and personalize the provided services. In the particular case of Search, this identification will trigger result personalization. Even if the user is not logged, best effort personalization of search results can be performed based on the recent browser activity.
\item \textsc{ssid}: this secured session cookie provides access to services that contains user data and personal information like \emph{Gmail, Google Calendar, Google Contacts,} etc.
\end{itemize}
It is our understanding that unlike the \textsc{ssid} cookie, the \textsc{sid} cookie just has an identification purpose and  can not be used to authenticate a user. On the other hand, the \textsc{ssid} cookie is sent only over encrypted connections and is required for services providing access to users' data and personal information. Google is thus implementing a two-level cookie based access policy; the first level only requires user identification while accessing to the second level assumes user authentication.\\
\tab In our study, we focus on the \textsc{sid} cookie and the information leakage that results of subsequent service personalization.

\begin{table*}[ht]
\small
\begin{tabular}[width=\textwidth]{l|l l l|l}

\multicolumn{2}{c}{\textbf{URL not available in HTTPS}} & &\multicolumn{2}{c}{\textbf{URL not redirected to HHTPS}} \\ 
\cline {1-2}  \cline {4-5}
\multicolumn{1}{c|}{Specialties} & \multicolumn{1}{|c}{Services} & &\multicolumn{1}{c|}{Specialties} & \multicolumn{1}{|c}{Services}  \\
\cline {1-2}  \cline {4-5}

google.com/blogsearch & picasa.google.com& & google.com/sitesearch &  investor.google.com \\
google.com/dirhp & maps.google.com & & google.com/transparencyreport & desktop.google.com \\
google.com/alerts & knol.google.com/k & & google.com/adsense/support & \\
google.com/mobile & webaccelerator.google.com & & google.com/insights/search & \\ 
google.com/nexus & sketchup.google.com & & google.com/prdhp & \\
google.com/analytics & books.google.com & &google.com/appsstatus & \\
google.com/postini & video.google.com & &google.com/chromebook & \\ 
google.com/chrome & scholar.google.com & &google.com/patents & \\ 

google.com/wallet & gears.google.com \\
google.com/ads/preference & \\
google.com/baraza &  \\
google.com/imghp &  \\

\end{tabular}
\normalsize
\caption{List of URLs that require the victim to send cleartext her cookie}
\label{tb:tablename}
\end{table*}

\subsection{Setting up the Attack}
\label{cap cookie}
The \textsc{sid} cookie is valid over the entire (\url{*.google.com}) domain, so it is sent to every Web application hosted on Google (\url{*.google.com}). Some services, such as \emph{Gmail}, are only available via a secured \textsc{https} connection, whereas others can be accessed through clear connection (\textsc{http}) that is the case of \emph{Google Search} and other services listed in Table \ref{tb:tablename}.\\
\tab The \textsc{sid} cookie is sent every time a request to a service under the (\url{google.com}) domain is sent, even when the queried page cannot be personalized (e.g:  privacy policy, terms of services). Because this cookie is sent to many \textsc{url}s, it is enough for one of these \textsc{url}s not to be accessible through \textsc{https} to be able to compromise it. 
\subsubsection{Bypassing HTTPS enforcement policy}
HTTPS-Everywhere \cite{HTTPS-Everywhere} is a browser extension that, when available, redirects a user to the secured version of the requested service to prevent traffic interception. Unfortunately this approach suffers from several drawbacks:
\begin{itemize}
	\item First, not every service is yet available through \textsc{https}. For instance, \emph{Google Alerts} remains only accessible through \textsc{http}.
	\item Second, the list of services available through \textsc{https} has to be maintained. Some services are already available in \textsc{https} but not yet redirected. A list of such services is reported in Table \ref{tb:tablename}.
  \item Finally, some services are redirected while not yet available. As a result these services can not be reached by HTTPS-Everywhere users. 
\end{itemize}
Due to these flaws, even HTTPS-Everywhere users could be redirected to a URL where they would have to send their \textsc{sid} cookies in cleartext.
\subsubsection{Intercepting the \textsc{sid} cookie}
Whenever a user accesses to one of the listed URL, her vulnerable \textsc{sid} cookie is exposed and so is her personal information. The objective of the attacker is to force the victim to exchange this cookie with unsecured services over a shared network, and so easily capture the cookie. Here are some examples.
\begin{itemize} 
\item The attacker can setup an open access point with a name (\textsc{ssid}) corresponding to a local hot-spot (\emph{e.g.} fast-food restaurant name) and include in the welcome page an hidden iframe pointing to an unsecured Google service. When connected, the browser will send to the rogue access point the valid \textsc{sid} cookie in clear text.
\item In an open wireless network, the attacker could also spoof the access point's physical address (\textsc{mac}) and respond to a victim's \textsc{http} \textsc{get} request by any Web page including the hidden iframe just like in the previous case.
\end{itemize}
\subsubsection{Googling for Cookies}
The simpler solution to find SID cookies is to search them. Typing the query \emph{``pref=id= sid= google'}'  in Google provides a list of pages where people published captured HTTP traffic, including SID cookies. Using the ``Past Month'' search filter increases the chances of retrieving valid SID cookies. Not all these results contain full SID cookies and some of the listed SID cookies may have already expired, but this simple search should already provide many valid cookies.

\section{Personalized search attack}
\label{retrieving data}
The search results provided to users who enabled \emph{Google Web Search History} are personalized and colored based on their previous interactions. One advantage of this feature is the ability to see the websites a user previously visited in the Google's search results. Furthermore, frequently visited pages are more likely to be high ranked. In 2009, Google also started to consider social network indicators as part of the inputs used to improve the search algorithm. ``Social Search'' now up-ranks results that user's contacts shared publicly via \emph{Google Buzz}, \emph{Twitter} and other social networks.\\
\tab The personalized search algorithm is based on private data held in the user's account. As such, it can be considered as a controlled information leakage. In \cite{Castelluccia:2010:PID:1881151.1881154}, authors suggest that this information leakage can be used to know some of the results a user clicked on. This flaw has not yet been fixed as considered hard to exploit: an attacker must know what the victim searched for and then compare the personalized and un-personalized results. Similarly, the visited-link coloration feature/vulnerability \cite{Castelluccia:2010:PID:1881151.1881154} was not addressed as considered innocuous regarding the benefits it offered.\\
\tab However, since these flaws have been reported, Google introduced new features that could improve the efficiency of attacks based on this information leakage. This section shows how these features could be misused to compromise users' click history.

\subsection{Google Search filters}
In this section we list the features that made the \emph{information leakage} attack more critical. These features --  available when clicking on `` Show search tools'' --  could be misused to significantly reduce the number of queries one has to issue to retrieve the previously clicked search results (and so harder to detect).

\subsubsection{ Visited Results } 
Collecting a user's visited links via random regular searches may take a very long time as the result pages mix visited and unvisited hyperlinks. However, an update of the Google Search interface introduced new filters. For instance, the ``Visited'' one filters out unvisited results \footnote{To visit the result page containing only visited links, one can directly go to http://www.google.com/search?q=.com\&tbo=1\&tbs=whv:1}. By enabling this filter, the information leakage becomes critical: only visited links remain listed as search results, partly disclosing the user's click history.
\subsubsection{Social}
With the ``Social'' filter, pages commented, twitted or shared via a social platform are displayed. The user's social network is built up according to his \emph{Gmail} contacts and possibly other social applications such as Twitter, Livejournal, etc (once linked to his Google account). This filter does not only provide the list of shared links, but also the connections that exist between the user and his social peers. Moreover, the user's Gmail contacts that should be treated as strictly private data is exposed and so could be used for blackmailing, phishing or spamming attacks. While we did not investigated how many contacts this attack may compromise, it appears that people who share a lot of information or belong to a large social network, are very likely to be listed in the ``Social'' search result page.\\
\tab In addition to the list of \emph{ Gmail} contact, \emph{ Google+} users will also receive links that have been shared by people in their Circles whether or not these contacts belong to exposed Circles. Consequently, information about the victim private Circles could also be leaked and retrieved through the ``Social'' filter.  
\subsection{Capturing the victim's data}
The sole prerequisite to run this attack is to capture the \textsc{sid} cookie of a user with Web Search History enabled. \textsc{sid} cookies are not marked ``secured'' and so are usually sent cleartext. If the victim uses a secure connection (\textsc{https}), the attacker could use the iframe injection described in Section \ref{google services and cookies} to force her to send the \textsc{sid} cookie cleartext. Once the cookie has been intercepted, the attack is launched by opening a window on Google Search with the ``Visited'' filter enabled. \\
\tab To start the attack, we use a list of 15 terms composed of domain extension (\url{.com}, \url{.net}, \url{.org},\url{.us},\url{.edu},\url{.fr},\url{.co.}), words and acronyms likely to appear in \textsc{url}s (\url{.jsp}, \url{.asp}, \url{php}, \url{html}, \url{index}, \url{www}) and popular websites (\url{google}, \url{facebook}). \\
To maximize the chance of finding all the visited links, the program should not only parse the first page of results, but browse all of them. We implemented a prototype that parses the result page and then clicks on ``Next'' to display and parse the following and so on until the last page. With \emph{Google Instant} disabled -- the attacker can set his Google \textsc{pref} cookie to display 100 results per page instead of 10. Because the \textsc{pref} cookie is browser specific -- and not linked to the currently used Google Account -- the attacker search preferences won't be mirrored on the victim browser. The list of retrievable clicked URLs can therefore be browsed very quickly. However, this attack is destructive; as the program interacts with the result pages, queries entered during the attack will appear in the victim's Web Search History.\\
\tab We conducted an experiment we analyze in section \ref{implem} and which shows that on average, 40\% of the click history can be retrieved. For Google users who search only occasionally from their account, up to 80\% of the click history could be retrieved using this method.

\section{Implementation and Evaluation}
\label{implem}

In order to validate the applicability of our attack and to illustrate its easy deployment, we implemented a proof of concept as a Firefox extension and another extension to measure the number of visited links that were retrievable. Both tools are available online (see \url{http://unsearcher.org/sid-test}) and we describe them in this section. We then detail and analyze our experiment results.\\
A chrome extension has also been developped to quickly set the SID cookie for the \url{google.com} domain (this extension is also available online \url{http://unsearcher.org})

\subsection{Extending Firesheep}
In October 2010, the Firefox extension Firesheep \cite{Firesheep} was released and emphasized the simplicity of session hijacking. This extension monitors network interfaces to capture cookies corresponding to sessions established on popular Web Service websites like Facebook, Google and Twitter. Once a cookie is captured, the extension provides an access to the account related to the hijacked session. \\
\tab We extended Firesheep to implement our information leakage attack. Thanks to the Firehseep modularity, we easily added a module that performs the attack on the session hijacked by the original code.\\
\tab As a result, when a Google \textsc{sid} cookie is captured, the account name appears in the Firesheep sidebar. Double clicking on it starts the attack; double clicking again displays the retrieved list of visited links. 

\subsection{Measurement methodology}
\tab We asked ten users to run the experiment on their Google accounts, we provided them with an extension that  extracts from a user Web Search History the clicks recorded since 1st January 2011 and then issues some queries from the user's account with the ``Visited'' filter activated. The extension --- developed for this experiment --- and the corresponding instructions have been publicly released\footnote{The extension is available at \url{http://unsearcher.org/Test\%20Flaw/ad@monitor.xpi}} to let users evaluate which portion of their click history can be exposed. Notice that, in order to preserve privacy of our testers, we asked them to send us only the ratio of clicks the attack was able to retrieve.\\
\tab It is worth noticing that, among the 10 volunteers who all had Web Search History enabled, 6 were not aware that the service existed and never used it (on purpose). It has been estimated that 50\% of Google accounts have Web Search History enabled \cite{Castelluccia:2010:PID:1881151.1881154}, but many may not be aware that the service is enabled, as the opt-out option has not always been obvious \cite{Sullivan}.
\begin{table*}
\begin{tabular}{c| c c c c c c c c c c|c} \hline
USER&U1&U2&U3&U4&U5&U6&U7&U8&U9&U10&Avg\\ \hline \hline
Links (since Januray 2011) &88&111&211&426&625&812&1148&1340&2521&3059&1034\\ \hline
Found (since Januray 2011)  & 72 & 69 & 145 & 133 & 199 & 320 & 372 & 429 & 467 & 351 & 256\\ \hline
Query 1 & 66\% & 15\%  & 41\%  & 4\% & 16\%	 & 12\%	 &  23\%	 &  21\%	 & 12\%	 &  07\%	 &  22\%\\ 
Query 2 & 68\% & 50\%  & 60\%  & 26\% & 19\%	 & 23\%	 &  28\%	 &  26\%	 & 16\%	 &  09\%	 &  33\%\\ 
Query 4 & 72\% & 62\% & 66\%  & 29\% & 24\% & 34\% & 31\% & 29\% & 17\% & 10\% & 37\% \\ 
Query 15 & 82\% & 62\% & 69\% & 31\% & 32\% & 39\% & 32\% & 32\% & 19\% & 11\% & 41\% \\ \hline
Found   & 338 & 69 & 359 & 137 & 315 & 541 & 425 & 628 & 644 & 640 & 410\\ \hline
%Time & 2m22 & 1m40 & 2m40 & 1m48 & 2m13 & 4m27 & 3m36 & 4m14 & 3m48 & 4m & 3m06\\ \hline \hline
%(Silent exp) Links Total&NA &NA&NA&NA&NA&NA&639&NA&2850&2190&NA\\ \hline
%Links Retreived &NA &NA&NA&NA&NA&NA&391&NA&567&565&NA\\ \hline
%Ratio &  NA & NA & NA & NA & NA &  NA & 61\% & NA &  26\% & 20\% & NA \\ \hline
\end{tabular}
\caption{Summary of the experiment results}
\label{tb:tablename2}
\end{table*}
\normalsize

\subsection{Result Analysis}
\begin{figure}[ht]
\begin{center}
	\includegraphics[width=0.495\textwidth]{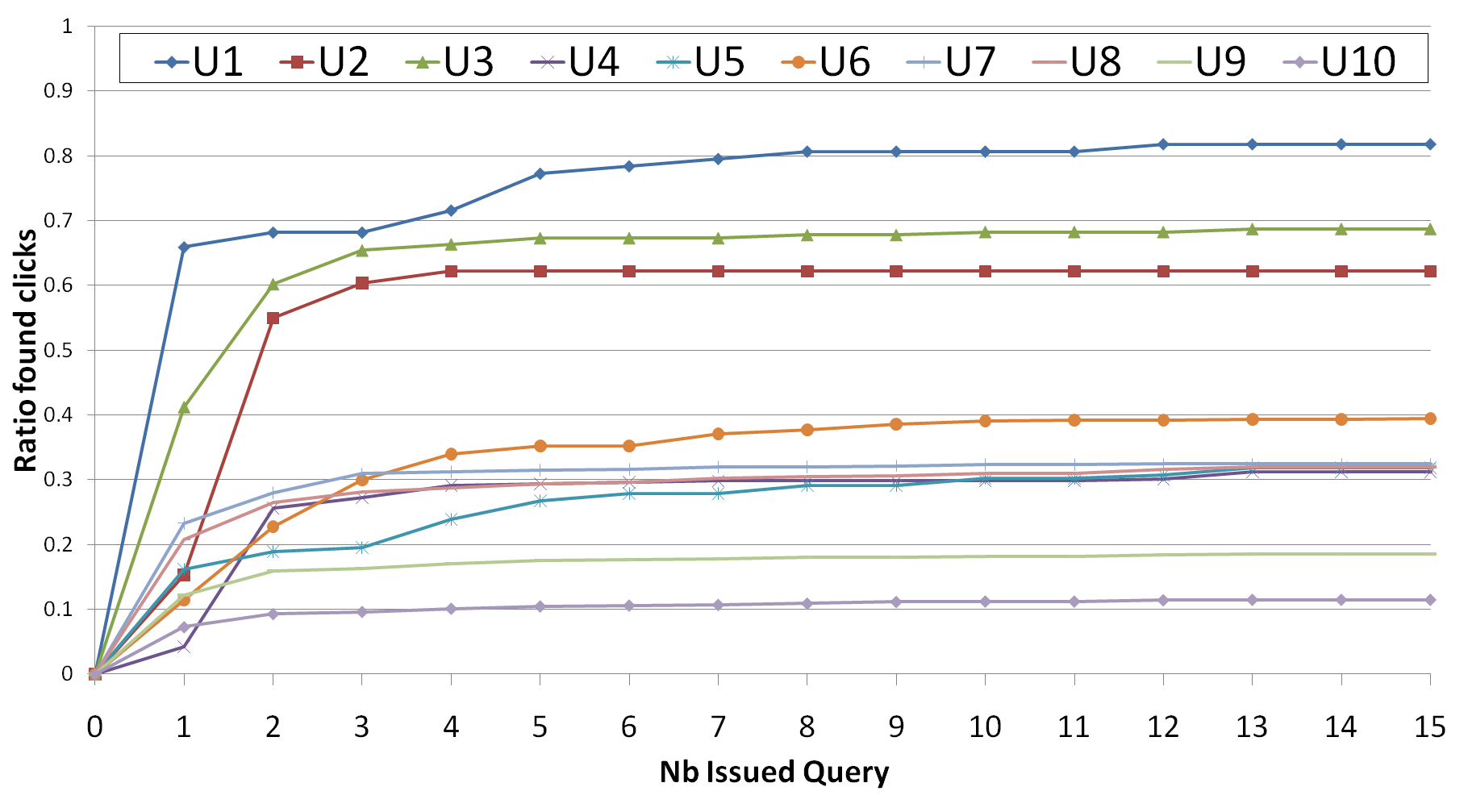}
	\caption{Destructive Approach: Result for the 10 users}
		\label{graph1}
\end{center}
\end{figure}

We summarize the results of our experiments in Table \ref{tb:tablename2}.We simulated the  attack on a set of ten volunteers who visited between 88 and 3059 search results between January and July 2011. For these users, between 72 and 467 links were retrieved counting for 82\% to 11\% of the links recorded in their Web Search History over the considered period. We also recorded the total number of links that the attack retrieved independently of the date they were visited. \\
\tab The attack provides similar results for the three users with the more visited links in their search history. For these three users who visited more links (\emph{Users 8, 9} and \emph{10}), a similar number of visited links were retrieved (between 628 and 644) although the number of clicks in the Web Search History varies from 1340 to 3059. \\ 
\tab Figure \ref{graph1} depicts the ratio of retrieved query as a function of the number of query submitted. We decided to use suffix and file extensions in our query list as they are very likely to appear in the URL of visited links and should return many search results. For instance, the query ``.com'' will return the list of visited links with the ``.com'' suffix. \\
\tab The objective is to recover a large portion of the victim's visited search results with the smallest set of search queries. Limiting the number of queries that will appear in the Web Search History reduces the risk that the victim detects the attack. A detected attack is likely to result in the purge of the victim's Web Search History and would prevent any further exploitation of the \textsc{sid} cookie by the attacker.\\
\tab Four queries --- with no associated visited search results --- are likely to remain unnoticed in victims' Web Search History and are enough to retrieve a large part of the clicked results. For all users, the attack submitted the 15 queries in less than 5 minutes.

\section{Countermeasure}
The \emph{information leakage} attack we described just require to read victims' \textsc{sid} cookies. If a user is not logged on her Google account when she accesses to Google services from a shared network, her \textsc{sid} cookie can not be compromised. A solution is therefore to sign out from Google accounts when connecting from a shared network or to use a \textsc{vpn} to encrypt the traffic and prevent cookie interception.\\
\tab From a user perspective a solution to prevent this second attack is to purge the Web Search History and to disable temporarily this feature. Such radical solution would definitely prevent the leakage of information about the user's search history but would not prevent a list of \emph{Gmail} and \emph{Google+} contacts to be exposed by ``Social Search''.\\
\tab Another solution is to disable the ``visited'' and ``social''  search filters when  a user is not visiting the secured version of Google. While this -- most likely --temporary solution does not fist the information leakage, it makes its exploitation more complicated and detectable.

\section{Conclusion}
\label{conclusion}
We presented an information leakage attack that leverages Google two-level cookie based access policy. We described an information leakage attack, implemented a proof of concept and evaluated the number of links visited over the last six months that could be exposed. Both issues should soon be fixed and we describe measures users could take to protect their search history from this kind of attack.\\
\tab Nevertheless, some issues can not be addressed by users and require a modification of Google's cookie policy. As Google is taking steps to include social indicators in result personalization, user's social network could soon be exposed. 

\bibliographystyle{plain}
\bibliography{WPES}

\end{document}